\documentclass[aps,pra,groupedaddress,showpacs,twocolumn]{revtex4-1}
\usepackage{graphicx}
\usepackage{amsmath}
\usepackage{bbm}
\usepackage{xspace}
\usepackage{color} 
\usepackage{hyperref}

\usepackage[normalem]{ulem} 

 \definecolor{eacol}{cmyk}{0,0.81,0,0}

\definecolor{orange}{RGB}{252,77,6}
\definecolor{brown}{RGB}{200,127,50}
\definecolor{light}{gray}{0.70}

\newcommand{\tra}[2]{(\mbox{$#1$$\rightarrow$$#2$})}
\newcommand{\tls}{TLS}
\newcommand{\np}{{N_p}}

\newcommand{\ie}{i.\thinspace{}e.\@\xspace}
\newcommand{\eg}{e.\thinspace{}g.\@\xspace}

\newcommand{\DF}[2]{\frac{d #1}{d #2}}

\newcommand{\tol}{{\mathcal{T}}}
\newcommand{\nag}{{\phantom{\dagger}}}

\newcommand{\eq}[1]{Eq.\thinspace{}(\ref{#1})}
\newcommand{\eqq}[2]{Eqs.\thinspace{}(\ref{#1}) and (\ref{#2})}

\newcommand{\fig}[1]{Fig.\thinspace{}\ref{#1}}

\newcommand{\fc}[1]{({#1})}
\newcommand{\figc}[2]{Fig.\thinspace{}\ref{#1}\thinspace{}\fc{#2}}
\newcommand{\figcc}[3]{Fig.\thinspace{}\ref{#1}\thinspace{}\fc{#2} and \fc{#3}}
\newcommand{\Fig}[1]{Fig.\thinspace{}\ref{#1}}

\newcommand{\Tr}{\mbox{Tr}}

\def\bra#1{\mathinner{\langle{#1}|}}
\def\ket#1{\mathinner{|{#1}\rangle}}

\DeclareMathOperator{\spn}{span}

\DeclareMathOperator{\vet}{vec}
\DeclareMathOperator{\id}{{\mathbbm{1}}}
\DeclareMathOperator{\idh}{\hat{\mathbbm{1}}}

\begin{document}



\title{Emission characteristics of laser-driven dissipative coupled-cavity systems} 


\author{Michael Knap}
\email[]{michael.knap@tugraz.at}
\author{Enrico Arrigoni}
\author{Wolfgang von der Linden}
\affiliation{Institute of Theoretical and Computational Physics, Graz University of Technology, 8010 Graz, Austria}
\author{Jared H. Cole}
\affiliation{Institut f{\"u}r Theoretische Festk{\"o}rperphysik and DFG-Center for Functional Nanostructures (CFN), Karlsruhe Institute of Technology, 76128 Karlsruhe, Germany}


\date{\today}

\begin{abstract}

We consider a laser-driven and dissipative system of two coupled cavities with Jaynes-Cummings nonlinearity. In particular, we investigate both incoherent and coherent laser driving, corresponding to different experimental situations. We employ Arnoldi time evolution as a numerical tool to solve exactly the many-body master equation describing the non-equilibrium quantum system.  We evaluate the fluorescence spectrum and the spectrum of the second-order correlation function of the emitted light field. Finally, we relate the measured spectra of the dissipative quantum system to excitations of the corresponding non-dissipative quantum system. Our results demonstrate how to interpret spectra obtained from dissipative quantum systems and specify what information is contained therein.

\end{abstract}

\pacs{05.70.Ln,71.36.+c,42.50.Pq,64.70.Tg}

\maketitle


Recent theoretical advances in quantum electrodynamics (QED) proposed that arrays of coupled cavities, containing some form of optical nonlinearity, 
allow for a realization of a phase with strongly correlated excitations and photons
\cite{hartmann_quantum_2008,tomadin_many-body_2010}. The optical nonlinearity might be achieved for instance by coupling the cavity photons to a two-level system ({\tls}) in the form of an atom or an atom-like structure. Individual accessibility and high tunability of the parameters make coupled cavity systems ideal candidates for quantum simulators and, probably even more intriguingly, for applications in the field of quantum information processing. Initial theoretical work \cite{hartmann_strongly_2006, greentree_quantum_2006, angelakis_photon-blockade-induced_2007} indicated a quantum phase transition of light from a localized Mott to a delocalized superfluid phase. Subsequently, the quantum phase transition \cite{rossini_mott-insulating_2007,hartmann_strong_2007, rossini_photon_2008,makin_quantum_2008,irish_polaritonic_2008,aichhorn_quantum_2008,zhao_insulator_2008,
lei_quantum_2008,na_strongly_2008,koch_superfluid--mott-insulator_2009,schmidt_strong_2009,pippan_excitation_2009,knap_jcl_2010,schmidt_excitations_2010,knap_quantum_2010} and the polaritonic nature of the constituent particles \cite{knap_jcl_2010,knap_quantum_2010} have been investigated in great detail. In single-cavity QED experiments, an astounding high level of control has already been achieved, allowing the observation of numerous fascinating phenomena based on the nonlinearities at the single photon level \cite{wallraff_strong_2004,yoshie_vacuum_2004,birnbaum_photon_2005,schuster_resolving_2007,hennessy_quantum_2007,sillanpaa_coherent_2007,majer_coupling_2007,srinivasan_linear_2007,schuster_nonlinear_2008,hofheinz_generation_2008,fink_climbing_2008,faraon_coherent_2008,fink_dressed_2009,bishop_nonlinear_2009,niemczyk_circuit_2010,johnson_quantum_2010,alton_strong_2010}. Importantly, in realistic experiments, these cavity QED systems are out of equilibrium, as they are driven by external lasers and are inherently susceptible to photon loss. 
Therefore the next step, having in view a scalable system of coupled cavities, is to understand the rich physics promised by a laser-driven and dissipative system of such coupled cavities, when out of equilibrium. 
\begin{figure}
        \centering
        \includegraphics[width=0.45\textwidth]{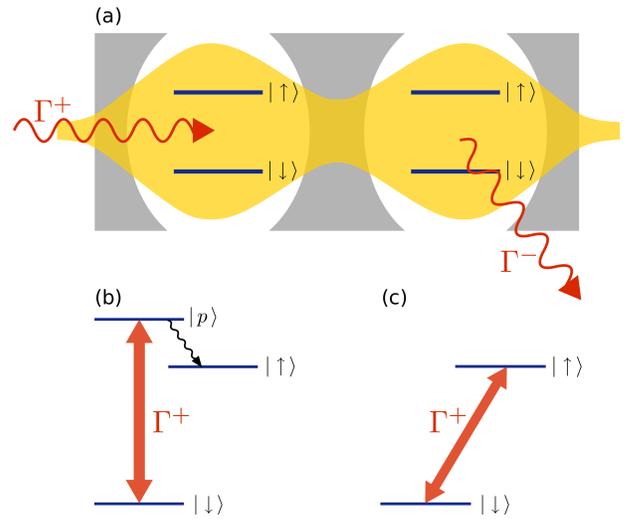}
        \caption{(Color online) Sketch  of two laser-driven and dissipative coupled cavities \fc{a}. Each cavity confines photons and contains a two-level system. The coupling of the cavities arises due to the finite spatial overlap of their photonic wave functions as indicated by the yellow area. The first cavity is driven either incoherently \fc{b} or coherently \fc{c}, with rate $\Gamma^+$, see main text for details. The photon dissipation out of the second cavity occurs at rate $\Gamma^-$.  }
        \label{fig:jch}
\end{figure}

Only very recently the theoretical investigation of such laser-driven and dissipative systems of coupled cavities has begun. The main research focus has been on the statistics of photons emitted from cavities with Kerr nonlinearity, which are coherently driven by an external laser source \cite{gerace_quantum-optical_2009,carusotto_fermionized_2009,hartmann_polariton_2010,liew_single_2010,ferretti_photon_2010,leib_bose-hubbard_2010,bamba_origin_2010,didier_detecting_2010}. {Steady-state entanglement of three coherently driven and dissipative cavities has been studied as well \cite{angelakis_coherent_2010}.} The time evolution of the population imbalance between two coherently driven cavities with Jaynes-Cummings (JC) nonlinearity \cite{jaynes_comparison_1963} has been investigated in Ref.~\onlinecite{schmidt_non-equilibrium_2010}. In addition, the non-equilibrium quantum phase transition of a dissipative and coherently driven cavity system with Kerr nonlinearity has been analyzed using a mean field approach \cite{tomadin_signatures_2010}. {Finally, investigations, within a slightly different context, on the non-equilibrium Bose-Hubbard model can be found in Refs.~\onlinecite{diehl_quantum_2008,diehl_dynamical_2010,tomadin_nonequilibrium_2010,pichler_non-equilibrium_2010}.}
 
In this paper, we investigate excitations of laser-driven and dissipative coupled cavities with JC nonlinearity.  To focus on the physical effects of interest, we introduce a `minimal model' consisting of two coupled cavities with a drive term applied to one two-level system and loss via photon emission from the other, see sketch in \figc{fig:jch}{a}. 
The JC nonlinearity accurately describes the interaction of cavity QED systems both in the quantum (few photon) limit as well as in the semiclassical (many photon) limit \cite{schmidt_non-equilibrium_2010}. In addition, it allows for an investigation of detuning effects between the cavity photons and the {\tls} present within each cavity. 

In our model, we consider either incoherent driving or coherent driving on the {\tls}, corresponding to two distinct experimental situations. As a technical point, we introduce the Arnoldi time evolution method in order to solve numerically exact the time dependence of the non-equilibrium system. Specifically, we evaluate the fluorescence spectrum and the spectrum of the second-order correlation function for the {\tls}.  This corresponds to an experimental situation where the {\tls} has a non-negligible loss rate into the non-cavity or `transverse' modes.  We specifically consider emission  into such non-cavity modes as they are both conceptually and experimentally easier to unambiguously distinguish from the emission stemming from cavity loss itself.

The main goal of this work is to relate spectra measured for the laser-driven and dissipative quantum system to the excitation energies of the non-dissipative quantum system. We show that the fluorescence spectrum probes excitations between two adjacent particle number sectors, whereas 
the spectrum of the second-order correlation function probes density-density excitations within one particle number sector. The second-order correlation function of the transverse field emitted from cavity systems with a JC nonlinearity exhibits qualitatively different properties when compared to systems with a Kerr nonlinearity, as the emitted photons are always antibunched.

The remainder of this paper is organized as follows. In Sec.~\ref{sec:model} we introduce our model in detail, which describes a  laser-driven and dissipative system of two coupled cavities with JC nonlinearity. In Sec.~\ref{sec:corr} we discuss the evaluation of first- and second-order steady-state correlation functions and spectra using the quantum regression theorem \cite{carmichael_statistical_2002}. The Arnoldi time evolution, which is the numerical method used to calculate the steady-state correlation functions is introduced in Sec.~\ref{sec:arnoldi}. 
Section~\ref{sec:inchoherent} is devoted to incoherent driving as a pump mechanism. In particular, single-particle excitation spectra and second-order correlation functions of the {\tls} situated in the second cavity are evaluated and discussed. Coherent driving is investigated in Sec.~\ref{sec:coherent}. Finally, we discuss and conclude our findings in Sec.~\ref{sec:conclusion}.

\section{\label{sec:model} Two laser-driven and dissipative coupled cavities}
In this paper, we investigate a laser-driven and dissipative system of two coupled cavities. A single cavity at site $i$ is modeled by the JC Hamiltonian~\cite{jaynes_comparison_1963}
\begin{equation}
 \hat{H}^{JC}_i = \omega_c \, a_i^\dagger \, a_i^\nag + \epsilon \, \sigma_i^+ \, \sigma_i^-+g( a^\nag_i\,\sigma_i^+ + a_i^{\dagger}\,\sigma_i^- ) \;\mbox{,}
 \label{eq:jc}
\end{equation}
where we set $\hbar=1$ as we do throughout this paper. The operator $a_i^\dag$ ($a_i^\nag$) creates (destroys) a photon of frequency $\omega_c$ at lattice site $i$. The {\tls} is mathematically described by Pauli spin algebra, where $\sigma_i^+$ ($\sigma_i^-$) is the raising (lowering) operator. The lower energy level will be denoted by $\ket{\downarrow}$ and the upper energy level by $\ket{\uparrow}$. The energy spacing of the {\tls} is $\epsilon$, and $g$ is the dipole coupling strength between photons and the {\tls}. In the subsequent analysis and calculations we use $g$ as our unit of energy. When considering the JC model in the rotating frame of the photon frequency $\omega_c$, the only remaining parameter is the detuning $\Delta$ between the photons and the two-level excitations $\Delta \equiv \omega_c - \epsilon$.

The total Hamiltonian of two coupled cavities is given by
\begin{equation}
 \hat{H}  = -J (a_1^\dag \, a_2^\nag+a_1^\nag \, a_2^\dag) + \sum_{i=1}^2 \hat{H}_i^{JC}\;.
 \label{eq:jcl}
\end{equation}
The coupling of the cavities arises due to the finite overlap of their photonic wave functions, which determines the hopping strength $J$, see sketch in \figc{fig:jch}{a}.

In experiments cavity systems are driven by an external laser source and are susceptible to photon loss. Therefore we have to investigate an open quantum system, whose dynamics are recovered under certain approximations \cite{carmichael_statistical_2002,breuer_theory_2002,gardiner_quantum_2004} by a master equation for the density operator $\hat \rho$ 
\begin{equation}
 \DF{}{t}\hat{\rho} = \hat{\mathcal{L}} \hat \rho\;.
 \label{eq:meq}
\end{equation}
The superoperator $\hat{\mathcal{L}}$ acts on the density operator $\hat \rho$ as 
\begin{equation}
 \hat{\mathcal{L}}\hat \rho = - i [\hat H, \, \hat \rho] + \sum_j \Gamma_j (\hat L_j^\nag \hat \rho \hat L_j^\dag - \frac12 \lbrace \hat L_j^\dag  \hat L_j^\nag, \hat \rho\rbrace)\;.
 \label{eq:superop}
\end{equation}
The first term on the right hand side of \eq{eq:superop} accounts for the unitary time evolution and the second term for decoherence processes. The operator $\hat L_j^\nag$ is the Lindbladian corresponding to the $j$th decoherence channel with characteristic decoherence rate $\Gamma_j$. Square brackets indicate the commutator $[ \hat a , \hat b] = \hat a\,\hat b - \hat b\,\hat a$ and curly brackets the anticommutator $\lbrace \hat a , \hat b\rbrace = \hat a\,\hat b + \hat b\,\hat a$.

In our investigations, we consider two distinct kinds of laser driving corresponding to two different experimental situations. The first is \textit{incoherent driving}, which is sketched in \figc{fig:jch}{b} and takes advantage of a three level structure. An external laser drives the transition from $\ket{\downarrow}$ to $\ket{p}$ coherently. A part of the excitations in $\ket{p}$ decays to the metastable state $\ket{\uparrow}$ by spontaneous emission, which leads to occupation of the upper state of the {\tls} considered in the JC Hamiltonian, see \eq{eq:jc}. This driving can be treated by an incoherent pump term, leading to the Lindbladian $L_1=\sigma^+_1$ with {effective pump} rate $\Gamma_1=\Gamma^+$, where we assume that the first cavity is driven by the external laser as indicated in \figc{fig:jch}{a}.

The second kind of driving we consider is \textit{coherent  driving}, where the {\tls} is directly driven by a laser of frequency $\omega_L$, see \figc{fig:jch}{c}. This leads to an additional Hamiltonian term \cite{alsing_spontaneous_1991}
\[
 \hat H^{drv} = F e^{-i \omega_L t} \sigma^+_1 +  F^* e^{i \omega_L t} \sigma^-_1\;,
\]
which must be added to $\hat {H}$. When considering the driving strength $F=\Gamma^+$ to be real, which affects only a global phase, and writing the Hamiltonian in the rotating frame of the laser frequency $\omega_L$, $\hat H^{drv}$ reduces to
\begin{equation}
 \hat H^{drv} = \Gamma^+ (\sigma^+_1 +  \sigma^-_1)\;.
 \label{eq:coh}
\end{equation} 
The impact of the rotating frame on the total Hamiltonian $\hat H$ is that there exist two detunings, specifically, the detuning between the laser and the cavity photons $\delta_{\omega_c} = \omega_L -\omega_c $ and the detuning between the laser and the energy spacing of the {\tls} $\delta_{\epsilon} = \omega_L -\epsilon $.

In cavity and circuit QED experiments, many loss channels have to be considered for photon decay. Among them are cavity losses and losses arising from decay of the {\tls}s into non-cavity modes. Our goal is not to describe all the details of a certain experiment, but to investigate the simplest possible model, that covers the essential physics of laser-driven and dissipative cavity systems. Thus we consider only one loss channel coupling to the {\tls} of the second cavity, meaning we consider a Lindbladian loss term  $L_2=\sigma^-_2$ with rate $\Gamma_2=\Gamma^-$. A sketch of the two coupled cavities with driving and dissipation is shown in \figc{fig:jch}{a}.

\section{\label{sec:corr} Correlation functions}
The purpose of this paper is to analyze spectra and correlation functions of photons emitted from the {\tls} of the second cavity. We investigate the {\tls} correlation functions, assuming that emission of the {\tls} into non-cavity modes can be differentiated from ordinary cavity losses.  This process is best visualized as the emission of the {\tls} into modes that are transverse to the plane of the cavity confined photonic modes.

All quantities are evaluated in steady state $\rho_{ss}$, which is obtained by solving \eq{eq:meq} for $d\hat \rho/dt =0$, \ie, by solving $\hat {\mathcal{L}} \hat \rho=0$. From the experimental viewpoint observables evaluated in steady state, \ie, in the long time limit, are of direct relevance, as in most experiments there is enough time between turning on the lasers and performing the measurements for the system to reach equilibrium.

In particular, we calculate the fluorescence spectrum, which is obtained from the Fourier transformation of the first-order correlation function in the steady state limit
\begin{equation}
 S(\omega) = \int \frac{dt}{\sqrt{2\pi}} \,e^{i \omega t} \, \langle \sigma_2^+(t) \, \sigma_2^-(0) \rangle_{ss}\;,
 \label{eq:a}
\end{equation}
where $\langle \sigma_2^+(t) \, \sigma_2^-(0) \rangle_{ss} \equiv  \lim_{\tau \to \infty} \langle \sigma_2^+(\tau+t) \, \sigma_2^-(\tau) \rangle$. The correlation function can be conveniently evaluated using the quantum regression theorem \cite{carmichael_statistical_2002}. 
For $t>0$ we have
\begin{equation}
 \langle \sigma_2^+(t) \, \sigma_2^-(0) \rangle_{ss} = \Tr (\sigma_2^+ \, e^{\hat {\mathcal{L}} t} \, \sigma_2^-\, \rho_{ss})\,,
 \label{eq:g1}
\end{equation}
where $\exp[{\hat{\mathcal{L}} t}]$ is the formal solution of the master equation in \eq{eq:meq}. 

In addition to the fluorescence spectrum, we analyze the spectrum $G(\omega)$ of the second-order correlation function $g^{(2)}(t)$. The second-order correlation function for the {\tls} of the second cavity is defined as 
\begin{equation}
 g^{(2)}(t) \equiv \frac{\langle \sigma_2^+(0) \, \sigma_2^+(t) \, \sigma_2^-(t)\, \sigma_2^-(0) \rangle_{ss}}{ \langle \sigma_2^+(0) \, \sigma_2^-(0) \rangle_{ss} \langle \sigma_2^+(t) \, \sigma_2^-(t) \rangle_{ss}}\;.
 \label{eq:g2}
\end{equation}
We again employ the quantum regression theorem and obtain for $t>0$
\begin{equation}
 g^{(2)}(t) \equiv \frac{\Tr [ \sigma_2^+ \, \sigma_2^- \,e^{\hat{\mathcal{L}} t} \, \sigma_2^-\,\rho_{ss}\, \sigma_2^+ ]}{(\Tr [\sigma_2^+ \, \sigma_2^- \,\rho_{ss}] )^2}\;.
 \label{eq:g2qrt}
\end{equation}
The second-order correlation function $g^{(2)}(t)$ provides information about the statistics of the photons. Specifically, it is the probability to detect two photons separated by a time delay $t$. 
In the case of a coherent light field $g^{(2)}(t)=1$ for all $t$ indicating a completely random sequence of photon pulses leading to a Poissonian distribution of time intervals between the pulses. 

However, when 
{$g^{(2)}(0)<1$} the emitted light exhibits distinct quantum mechanical features, \ie,\ antibunching \cite{paul_photon_1982, carmichael_statistical_2002}. Antibunched light leads to a sub-Poissonian distribution of time intervals between the emitted photons. The light emitted from the {\tls} considered here is perfectly antibunched, due to the nature of a {\tls}, since $(\sigma_2^+)^2 = (\sigma_2^-)^2 = 0$ and thus we have $g^{(2)}(0)=0$, see \eq{eq:g2}.
We introduce the Fourier transformation of the second-order correlation function as 
\begin{equation}
 G(\omega) = \int \frac{dt}{\sqrt{2\pi}} \,e^{i \omega t} \, g^{(2)}(t) \;,
 \label{eq:s}
\end{equation}
which, as we show below, contains information about the density-density correlations.

\section{\label{sec:arnoldi} Arnoldi time evolution}

In this section, we introduce the Arnoldi time evolution for open quantum systems, to obtain the full quantum mechanical solution of the master equation given in \eq{eq:meq}.
To this end, we rewrite the master equation as a matrix equation of the form 
\begin{equation}
 \DF{}{t}\vec \rho = \mathcal{L} \vec \rho\;,
 \label{eq:meqMat}
\end{equation}
where $\vec \rho \equiv \vet (\hat \rho)$ is a column-ordered vector representation of the density matrix and $\mathcal{L}$ is the superoperator in matrix form, which includes effects of both the unitary time evolution and the decoherence processes. Using tensor product properties \cite{fuller_mathematics_1992, barnett_matrices_1997}
the following identity holds
\begin{equation}
 \vet ( \hat A \, \hat X \,\hat B) = ( A \otimes  B^T) \vec X \;.
\label{eq:ve}
\end{equation}
Thus with judicious insertion of the identity operator $\idh$ into \eqq{eq:meq}{eq:superop} we have
\begin{align}
 \idh \DF{}{t} \hat \rho \idh &= -i (\hat H \hat \rho \idh - \idh \hat \rho \hat H) \nonumber\\
  &+ \sum_j \Gamma_j [\hat L_j^\nag \hat \rho \hat L_j^\dag - \frac12 ( \hat L_j^\dag  \hat L_j^\nag  \hat \rho \idh + \idh \hat \rho \hat L_j^\dag  \hat L_j^\nag   )]\;.
\end{align}
Applying the identity \eq{eq:ve} yields the wanted matrix representation of the master equation
\begin{align}
 \DF{}{t} \vec \rho &= -i [H \otimes \id - \id \otimes H^T ] \vec \rho \nonumber\\
  &+ \sum_j \Gamma_j [ L_j^\nag \otimes  L_j^*  - \frac12 \lbrace  L_j^\dag  L_j^\nag \otimes \id  + \id \otimes   (L_j^\dag   L_j^\nag)^T   \rbrace] \vec \rho\nonumber\\
  &= \mathcal{L} \vec \rho\;.
\end{align}
The superoperator matrix $\mathcal{L}$ is of size $N^2 \times N^2$ and $\vec \rho$ is a vector of size $N^2$, where $N$ is the Hilbert space dimension of the two cavity system.

The formal solution of \eq{eq:meqMat} is 
\begin{equation}
 \vec \rho(t) = e^{\mathcal{L}t} \vec \rho(0)\;.
 \label{eq:meqSol}
\end{equation}
In practice, the observables and thus the density operator have to be evaluated at multiple times separated by $\Delta t$ leading to the propagation prescription
\begin{equation}
 \vec \rho(t+\Delta t) = e^{\mathcal{L} \Delta t} \vec \rho(t)\;.
 \label{eq:meqSolDel}
\end{equation}

\begin{figure}
        \centering
        \includegraphics[width=0.4\textwidth]{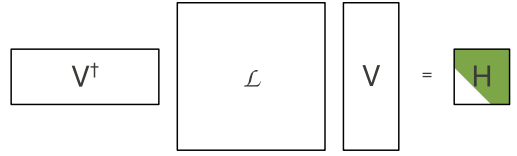}
        \caption{(Color online) Illustration of the projection of the superoperator matrix $\mathcal{L}$ onto an upper Hessenberg form $H$ using the Arnoldi basis $V$.  }
        \label{fig:arnoldi}
\end{figure}

We are now left with the calculation of the exponential $\exp[{\mathcal{L}\Delta t}]$, which amounts to diagonalizing the large, sparse, complex valued and non-Hermitian matrix $\mathcal{L}$. This can be achieved by the Arnoldi method \cite{arnoldi_principle_1951},
which is an iterative technique to approximate the eigenvalues of such matrices. In the following, we briefly sketch how the Arnoldi method works. Details can be found in Refs.~\onlinecite{saad_numerical_1995, saad_y_arnoldi_2000}
. This procedure is essentially based on constructing an orthonormal basis $V$ of the Krylov subspace 
\[ \mathcal{K}^m(\mathcal{L},\vec \rho) = \spn \lbrace \vec \rho, \, \mathcal{L} \vec \rho, \, \mathcal{L}^2\vec \rho, \, \ldots \, \mathcal{L}^{m-1}\vec \rho \rbrace\;.\]
In the following we will refer to $V$ as the Arnoldi basis. The orthonormal basis $V$ projects the superoperator matrix $\mathcal{L}$ onto an upper Hessenberg form $H$
\begin{equation}
 V^\dag \mathcal{L} V = H\;.
 \label{eq:arnoldi}
\end{equation}
An illustration of this projection is shown in \fig{fig:arnoldi}.
\begin{figure*}
        \centering
        \includegraphics[width=0.9\textwidth]{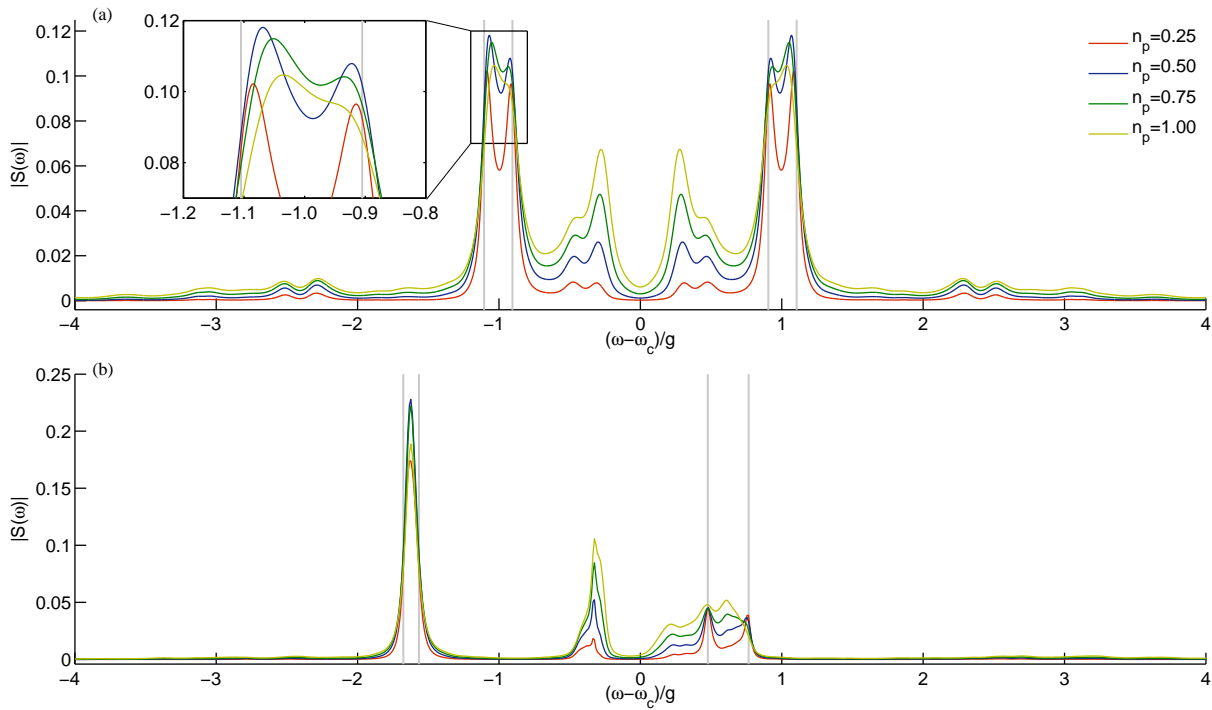}
        \caption{(Color online) Fluorescence spectra $S(\omega)$ of the laser-driven and dissipative two-cavity system for hopping strength $J/g=0.2$, loss rate $\Gamma^-/g=0.2$ and detuning \fc{a} $\Delta/g=0$ and \fc{b} $\Delta/g=1$, respectively. The driving strength $\Gamma^+$ is adjusted such that the steady-state particle density is $n_p=\lbrace 0.25,\,0.5,\,0.75,\,1.0 \rbrace$. In addition to the spectra, the light-gray (vertical) lines indicate single-particle excitations of the non-dissipative system from the $\np=0$ to $\np=1$ particle sector of the Hilbert space, which we refer to as \tra{0}{1}-transitions. These excitations turn out to yield the main contributions to the spectra. {The inset in panel \fc{a} magnifies the two-peak structure of $S(\omega)$ at $|\omega-\omega_c|\approx -g$ and demonstrates the shift of the resonances with varying driving strength.}  }
        \label{fig:sOfDensity}
\end{figure*}
The upper Hessenberg form $H$ is a small matrix of size $m\times m$ and can be diagonalized using standard methods
\begin{equation}
 H U = U D\;,
 \label{eq:hessen}
\end{equation}
where the matrix $U$ contains the eigenvectors of $H$ and the diagonal matrix $D$ its eigenvalues, which are Ritz approximate eigenvalues of $\mathcal{L}$ \cite{saad_numerical_1995}. 
Combining \eqq{eq:arnoldi}{eq:hessen} we have 
\[
 \mathcal{L} = V \, U \, D \, U^{-1} \, V^\dag\;.
\]
Since $V$ is an orthonormal basis, \ie, $V^\dag V = \id$, the formal solution given in \eq{eq:meqSolDel} can be expressed as
\[
 \vec \rho(t+\Delta t) = V \, U \, e^{D \Delta t} \, U^{-1} \, V^\dag \vec \rho(t)\,.
\]
This expression is evaluated straightforwardly as the matrix $D$ is diagonal. When choosing $\vec \rho(t)$ as initial vector of the Krylov subspace, $V^\dag \vec \rho(t)$ simplifies to $\vec e_1 = (1,\,0,\,0 \,\ldots)^T$ since all vectors of the Arnoldi basis $V$ are constructed to be orthogonal. With these considerations we have 
\begin{equation}
 \vec \rho(t+\Delta t) = V \, U \, e^{D \Delta t} \, U^{-1} \, \vec e_1\,.
 \label{eq:arnodliTime}
\end{equation}
The Arnoldi method is an iterative procedure and thus a convergence criteria has to be imposed. The propagation \eq{eq:arnodliTime} can be regarded from a slightly different viewpoint as an expansion in the Arnoldi basis vectors
\begin{equation}
 \vec \rho(t+\Delta t) = \sum_{i=1}^m c_i \vec v_i\;,
 \label{eq:arnoldiTimeExpansion}
\end{equation}
where $\vec v_i$ is the $i$th column of $V$ and $c_i$ is the $i$th component of the vector $\vec c \equiv U \, e^{D \Delta t} \, U^{-1} \, \vec e_1$. 
{A convenient convergence criteria for Krylov based time evolution \cite{park_unitary_1986} is 
\begin{equation}
|c_m|^2 < \tol\,,
\label{eq:conv}
\end{equation}
since an extension of the Krylov basis has no implications on the result within the order of the tolerance $\tol$. Importantly, the choice of both the time step $\Delta t$ and the size of the Arnoldi basis $m$ influences the convergence. In practice, the time step is chosen to be a trade off between the maximally reached time and the computer time (and memory) needed to perform one Arnoldi time step $\Delta t$, which of course depends on the Krylov basis size $m$.

Alternatively to employing the Arnoldi time evolution, \eq{eq:meqMat} can be solved deterministically by conventional integration schemes based on Runge-Kutta like approximations (for an application see \eg Ref.~\onlinecite{tomadin_nonequilibrium_2010}) or by Laplace transforming the differential equation and subsequently solving numerically a sparse system of equations for each frequency of interest \cite{mollow_power_1969,mollow_pure-state_1975}. The latter approach is particularly suitable for small systems, since it is a numerically exact approach and easy to apply. However, for medium sized or larger problems solving the sparse system of equations is not feasible any longer due to stringent memory requirements. Both conventional integration schemes and the Arnoldi time evolution can be applied to medium sized problems. Compared to conventional integration schemes,  the Arnoldi time evolution method needs slightly more memory since, depending on the chosen timestep $\Delta t$, approximately 10 to 20 Krylov basis vectors have to be stored. The advantages of the Arnoldi time evolution, however, are that it is very stable in the long time limit and that a stringent convergence criteria [see \eq{eq:conv}] is available to control the accuracy of the evolved density matrix. Alternatively, to the deterministic approaches, the master equation can be solved stochastically using quantum trajectories \cite{mollow_pure-state_1975,dum_monte_1992,breuer_theory_2002,gardiner_quantum_2004,carmichael_statistical_2009,pichler_non-equilibrium_2010}. 

In our calculations, we employ the Arnoldi time evolution to evaluate the time dependent correlation functions from the full quantum mechanical master equation. The initial state for the time evolution is the steady-state density matrix $\rho_{ss}$ appropriately multiplied by raising and lowering operators of the {\tls}, compare with \eqq{eq:g1}{eq:g2qrt}. We evaluate the steady-state density matrix $\rho_{ss}$ by solving for the null space of the superoperator, \ie, by solving $\mathcal L \vec \rho = \vec 0$, which, in the case of our model, has a unique solution for any physically applicable set of parameters. 

Technically, solving $\mathcal L \vec \rho = \vec 0$ is achieved by a shift-and-invert Arnoldi method, where a small shift $\varepsilon$ regularizes $\mathcal L$
and the invert mode guarantees convergence toward the eigenvalue closest to $\varepsilon$. For small enough shifts $\varepsilon$, the shift-and-invert Arnoldi method converges toward the wanted eigenvalue $0$ of the superoperator $\mathcal L$, and thus, the corresponding eigenvector is the steady-state density matrix in column order. 

\section{\label{sec:inchoherent} Incoherent driving}
In this section, we present results for the incoherently driven system of two coupled dissipative cavities. Specifically, we evaluate fluorescence spectra $S(\omega)$ and spectra $G(\omega)$ of the second-order correlation functions $g^{(2)}(t)$.

\subsection{\label{subsec:spectra} Fluorescence spectra}
In order to understand quantum systems it is important to track properties gained from measurements on the realistic, dissipative  system back to properties of the non-dissipative system. Here, we investigate fluorescence spectra of the dissipative quantum system and relate the resonance peaks of the spectra to excitations of the non-dissipative system. In \fig{fig:sOfDensity} we show the fluorescence spectrum, \ie, the Fourier transform of the first-order correlation function, 
{see \eq{eq:a}.}

We choose {a} small hopping strength $J/g=0.2$, corresponding to the regime of strong correlations, since the infinitely large, non-dissipative coupled-cavity system exhibits a Mott to superfluid quantum phase transition at this parameter regime for small filling factors \cite{greentree_quantum_2006,rossini_mott-insulating_2007, rossini_photon_2008,makin_quantum_2008,irish_polaritonic_2008,aichhorn_quantum_2008,zhao_insulator_2008,koch_superfluid--mott-insulator_2009,schmidt_strong_2009,knap_jcl_2010}. 
Thus most interesting physics is to be expected in this parameter region. In addition, this parameter regime is also of great interest for the two-cavity system, as can be seen from the particle-density diagram in the hopping strength and chemical potential plane evaluated in Ref.~\onlinecite{makin_quantum_2008}. In this regime of the hopping strength, regions of considerable extent, for varying chemical potential, exist for arbitrary filling factors. The loss rate is set to $\Gamma^-/g=0.2$ and the strength of the driving laser $\Gamma^+$ is adjusted such that the steady-state density is $n_p=\lbrace 0.25,\,0.5,\,0.75,\,1.0 \rbrace$. As throughout this work we use the dipole coupling strength $g$ as the unit of energy. In \figc{fig:sOfDensity}{a} the spectra are evaluated for zero detuning $\Delta/g=0$ whereas in \fc{b} a finite detuning of $\Delta/g=1$ is considered. 

We employ the Arnoldi time evolution described in Sec.~\ref{sec:arnoldi} to evaluate the time dependent correlation functions from the full quantum mechanical master equation. 
As a technical point, it is important to note that the time evolution can be performed only for positive time steps, since negative time steps would lead to numerical instabilities of the exponential in \eq{eq:arnodliTime}. However, negative times can be attained by observing that the first-order correlation function evaluated for a negative time $-t$ is simply the complex conjugate of the one for positive time $t$, see \eq{eq:g1}. The Hilbert space of the two-cavity system is infinitely large, as each cavity can be occupied by infinitely many photons. For the numerical investigation we set the maximal number of photons per cavity to $8$, which was found to result in convergence for the weak driving limit considered here. 

The analogy of the fluorescence spectra in condensed matter physics is the single-particle excitation spectrum, which probes the energy necessary to add (remove) a particle to (from) the system. The Hamiltonian 
of the non-dissipative two-cavity system can be arranged in disconnected sectors of a certain particle number $\np$. Due to the block diagonal form of the Hamiltonian the Schr{\"o}dinger equation can be solved for each sector $\np$ separately leading to
\[
 \hat H \ket{\psi_\nu^{\np}} = E_\nu^\np \ket{\psi_\nu^{\np}} \;,
\]
where $E_\nu^\np$ are the eigenenergies and $\ket{\psi_\nu^{\np}}$ are the corresponding eigenvectors. The single-particle excitation spectrum probes excitations from the Hilbert space sector with $\np$-particles to $(\np+1)$-particles, corresponding to the energy difference $\omega_{\mu\nu}^\np \equiv E_\mu^{\np+1}-E_\nu^\np$. In the weak driving limit, we infer that the excitations from the zero-particle sector $\np=0$ to the one-particle sector $\np=1$ are the most dominant. In the following, we will refer to these excitations as \tra{0}{1}-transitions. The non-dissipative system can be solved analytically for these Hilbert space sectors, leading to
{\allowdisplaybreaks
\begin{align}
 E_1^0&=0 \nonumber \\
 E_1^1 &=\omega_{c} - (J + \Delta)/2 - \sqrt{(J-\Delta)^2/4 + g^2} \nonumber \\
 E_2^1 &=\omega_{c} - (J + \Delta)/2 + \sqrt{(J-\Delta)^2/4 + g^2} \nonumber \\
 E_3^1 &= \omega_{c} + (J - \Delta)/2 + \sqrt{(J+\Delta)^2/4 + g^2} \nonumber \\
 E_4^1 &= \omega_{c} + (J - \Delta)/2 - \sqrt{(J+\Delta)^2/4 + g^2} \;. 
 \label{eq:en}
\end{align}
}

In \fig{fig:ladder} we compare schematically the energy ladder for zero detuning $\Delta/g=0$ of \fc{a} a single JC cavity, obtained from the solution of \eq{eq:jc}, and of \fc{b} two coupled cavities. 
\begin{figure}
        \centering
        \includegraphics[width=0.3\textwidth]{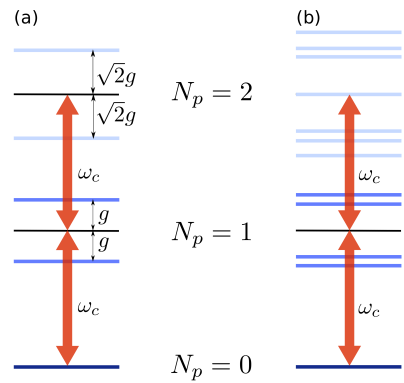}
        \caption{(Color online) Energy ladder for zero detuning $\Delta/g=0$ of \fc{a} a single cavity and of \fc{b} two coupled cavities. The number of particles in the respective Hilbert space sector is $\np$.}
        \label{fig:ladder}
\end{figure}
\begin{figure*}
        \centering
        \includegraphics[width=0.9\textwidth]{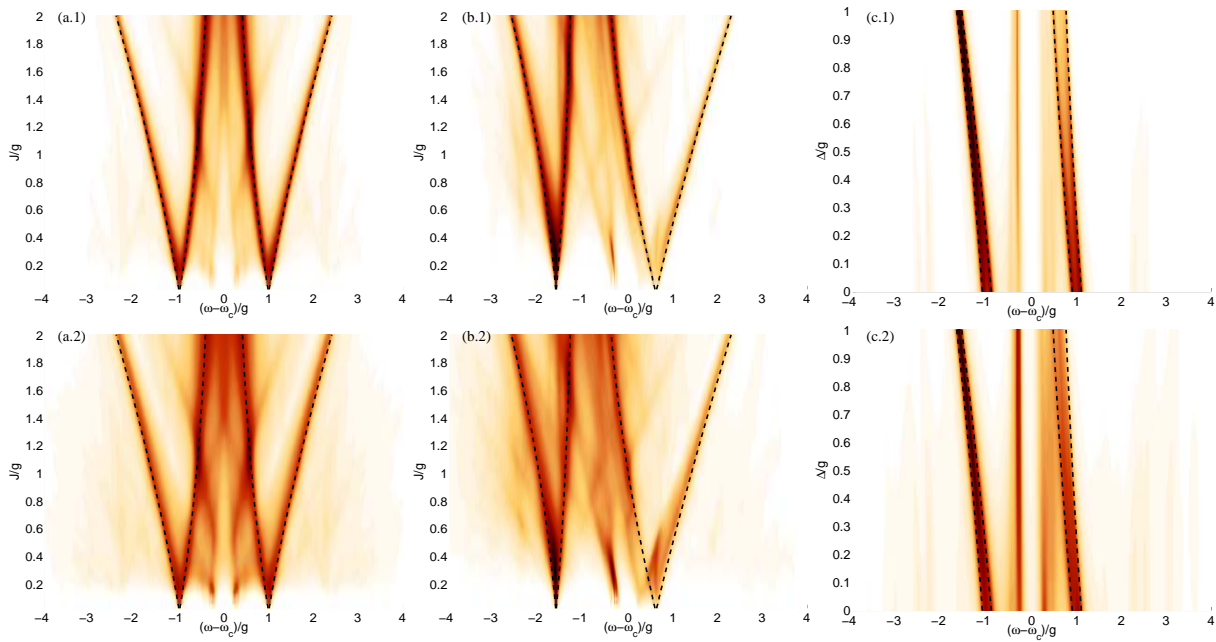}
        \caption{ (Color online)
        Density plot of the fluorescence spectra $S(\omega)$ for steady-state particle density $n_p=0.5$, top row, and $n_p=1.0$, bottom row. {The loss rate is set to $\Gamma^-/g=0.2$.}
        The first two columns depict the density as function of the hopping strength $J/g$ for two detuning parameters $\Delta/g=0$ [column \fc{a.\textasteriskcentered}] and $\Delta/g=1$ [column \fc{b.\textasteriskcentered}]. 
        The third column, labeled (c.\textasteriskcentered), shows the detuning-dependence of the fluorescence spectra for $J/g=0.2$. In all panels dark dashed lines indicate \tra{0}{1}-transitions of the non-dissipative system.}
        \label{fig:sfRec}
\end{figure*}
Most important for our analysis within the weak driving limit are the low energy excitations. {For zero coupling $J/g=0$ the eigenenergies of the $\np=1$ particle number sector  $\omega_c\pm g$ are two-fold degenerate, see \eq{eq:en}, and apparently correspond to the $\np=1$ eigenenergies of a single JC cavity. A finite coupling strength $J/g$, however, removes the degeneracy and leads to the four peak structure observed in  \fig{fig:sOfDensity}. }
The excitations $\omega_{\mu 1}^0 = E_\mu^{1}-E_1^0 $ obtained from \eq{eq:en}, correspond to the energy differences between these four levels and the ground state energy. In \fig{fig:sOfDensity} these energy differences are indicated by light-gray (vertical) lines and match as expected the main four resonances, present in the spectrum  of the dissipative quantum system. {Importantly, for increasing driving strength the resonances of the open quantum system are slightly shifted as demonstrated in the inset of \figc{fig:sOfDensity}{a}.} 

\Fig{fig:sfRec} shows density plots of the fluorescence spectra.
The top and bottom rows depict results for steady-state particle densities $n_p=0.5$ 
and $n_{p}=1$, respectively.
The first two columns show the spectra as function of the hopping strength $J/g$ for two different
detuning parameters $\Delta/g=0$ [column \fc{a.\textasteriskcentered}] and $\Delta/g=1$
 [column \fc{b.\textasteriskcentered}].
The third column, labeled (c.\textasteriskcentered), shows $S(\omega)$ for $J/g=0.2$ as a function of the detuning $\Delta$. {In all plots we set the loss rate to $\Gamma^-/g=0.2$.} The \tra{0}{1}-transitions of the non-dissipative system are indicated by dark dashed lines.
 In the case of weak driving the main contributions to the spectra of the dissipative two-cavity system correspond to these excitations.
However, for increasing driving strength, comparing the top and bottom rows of \fig{fig:sfRec}, excitations from higher particle number sectors become 
increasingly important.
In the spectra shown here, the main contributions to the additional peaks stem from the \tra{1}{2}-transitions.
Interestingly, for increasing driving strength the weight of the \tra{0}{1}-transitions stays almost constant, whereas the weight of the \tra{1}{2}-transitions increases, in the regime considered here, since the $\np=1$ particle number sector has to be occupied before higher excitations can take place. When comparing 
\figcc{fig:sfRec}{c.1}{c.2}
 it can be observed that  the detuning mainly affects the excitation energies of the \tra{0}{1}-transitions. The excitation energies of the \tra{1}{2}-transitions remain almost unaltered, only the intensities of $S(\omega)$ are slightly modified.

The general conclusion of  \fig{fig:sfRec} is that 
all poles $\omega_{\mu\nu}^\np$ resulting from excitations from the sectors with $\np$ to $\np+1$ particles contribute to the fluorescence spectra and the strength of the poles depend on the occupation probability of states associated with the transition.

\subsection{Approximate semianalytic spectrum}
To gain more insight into the excitations we \textit{approximately} evaluate the spectrum in a semianalytic way from the steady-state density matrix $\rho_{ss}$. To this end, we assume that each resonance 
$\omega_\alpha \equiv \omega_{\mu\nu}^\np
= E^{N_{p}+1}_{\mu} - E^{N_{p}}_{\nu}
$, 
where $\alpha$ is a collective index containing $\mu$, $\nu$ and $\np$, of the spectrum can be described by a Lorentzian, which is of the form
\[
\tilde S_\alpha(\omega) = l_\alpha \frac{(\Gamma/2)^2}{(\omega-\omega_\alpha)^2 +(\Gamma/2)^2 }\,,
\]
where $l_\alpha$ is the strength of the individual peak and $\Gamma$ is its half-width. We evaluate the half-width $\Gamma$ by
\[
 \Gamma=\Gamma^- \frac{n_2^\sigma}{n_2^{ph}+n_2^\sigma}\;,
\]
where $\Gamma^-$ is the loss rate and $n_2^\sigma$ ($n_2^{ph}$) is the {\tls} (photon) steady-state density of the second cavity, where the emission spectrum is measured. Next, we have to find an approximate prescription to evaluate the individual strength $l_\alpha$ of each pole $\omega_\alpha$. 
Motivated by the Green's function theory for closed quantum systems \cite{fetter.walecka}, we assume that the contribution of each pole is proportional to
\[
 l_\alpha=|\bra{\psi^{\np+1}_\mu} \sigma^+_2 \ket{\psi^{\np}_\nu} \bra{\psi^{\np}_\nu} \sigma_2^- \rho_{ss} \ket{\psi^{\np+1}_\mu}|\;;
\]
compare with the first-order correlation function given in \eq{eq:g1}. Following these considerations a Fourier transform of the approximated correlation function would yield a resonance peak located at the assumed energy $\omega_\alpha$, which once again proves our choice of $l_\alpha$ to be reasonable.
The full spectrum is constructed by summing over all individual Lorentzians
\[
 \tilde S(\omega) = \sum_\alpha \tilde S_\alpha(\omega) \;.
\]
Results obtained from this semianalytic treatment are shown along with the exact numerical results in \fig{fig:sfAn} for \fc{a} zero detuning $\Delta/g=0$ and \fc{b}  detuning $\Delta/g=1$. 
\begin{figure}
        \centering
        \includegraphics[width=0.39\textwidth]{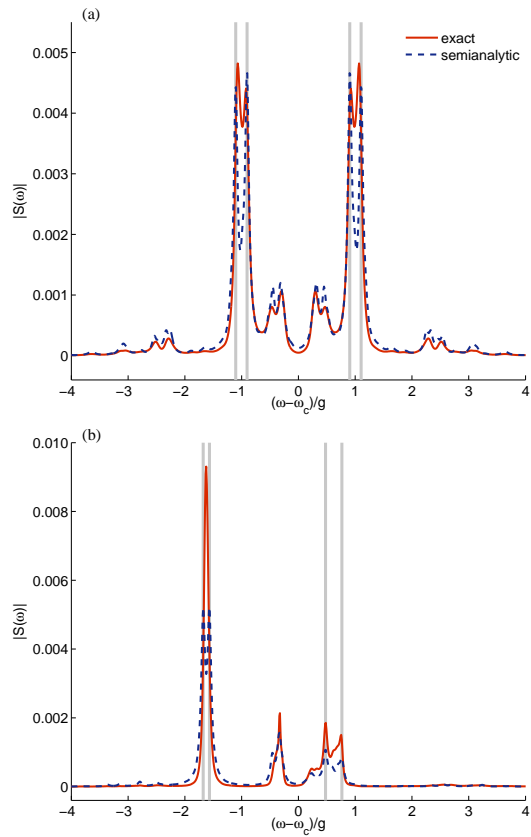}
        \caption{(Color online) Comparison between the exact spectrum, red, solid line, and the semianalytic spectrum, blue, dashed line. The hopping strength is $J/g=0.2$, the loss rate is $\Gamma^-/g=0.2$  and the driving strength $\Gamma^+$ is adjusted such that the steady-state particle density is $n_p=0.5$. Spectra are evaluated for detuning \fc{a} $\Delta/g=0$ and \fc{b} $\Delta/g=1$. Light-gray (vertical) lines indicate \tra01-transitions of the non-dissipative system. }
        \label{fig:sfAn}
\end{figure}
The hopping strength is $J/g=0.2$, the loss rate is $\Gamma^-/g=0.2$ and the driving strength $\Gamma^+$ is adjusted such that the steady-state particle density is $n_p=0.5$.  We normalize the semianalytic spectrum such that the area under the curve coincides with the area under the exact numerical spectrum. Overall the spectrum evaluated with this semianalytic technique fits the exact numerically evaluated spectrum rather well. In addition, to the four major resonances, peaks corresponding to excitations from Hilbert space sectors with $\np>0$ can be observed. 

\subsection{\label{subsec:g2} Second-order correlation functions}
Returning to our full numerical solution of the master equation, we are able to evaluate the time-dependence of the second-order correlation function $g^{(2)}(t)$ and its spectrum $G(\omega)$. We evaluate $g^{(2)}(t)$ for the {\tls} of the second cavity, since the transverse field emission can be conveniently separated from the emitted cavity photon field in experiments.
The second-order correlation function $g^{(2)}(t)$, see \eq{eq:g2}, is the probability of detecting two photons separated by a time delay $t$.
As required  for the second-order correlation function of a {\tls} $g^{(2)}(0)=0$, corresponding to a completely quantum mechanical, antibunched light source. 
This is in stark contrast to results obtained from a Hubbard model with Kerr nonlinearity, where a transition from bunched to antibunched light statistics can be observed as a function of the nonlinearity strength \cite{gerace_quantum-optical_2009,carusotto_fermionized_2009,hartmann_polariton_2010,liew_single_2010,ferretti_photon_2010,leib_bose-hubbard_2010,bamba_origin_2010,didier_detecting_2010}.
The second-order correlation function $g^{(2)}(t)$ is an oscillating function, which initially increases and eventually approaches its asymptotic value one, meaning that the two emitted photons are uncorrelated for large time delay $t$.
Such asymptotic behavior results in a pronounced peak in $G(\omega)$ at $\omega=0$. 
In \fig{fig:g2} we show the spectrum $G(\omega)$ for zero detuning $\Delta/g=0$, loss rate $\Gamma^-/g=0.2$, and driving strength $\Gamma^+$ adjusted such that the steady-state particle density is $n_p=\lbrace 0.5,\,1.0 \rbrace$. 
\begin{figure}
        \centering
        \includegraphics[width=0.39\textwidth]{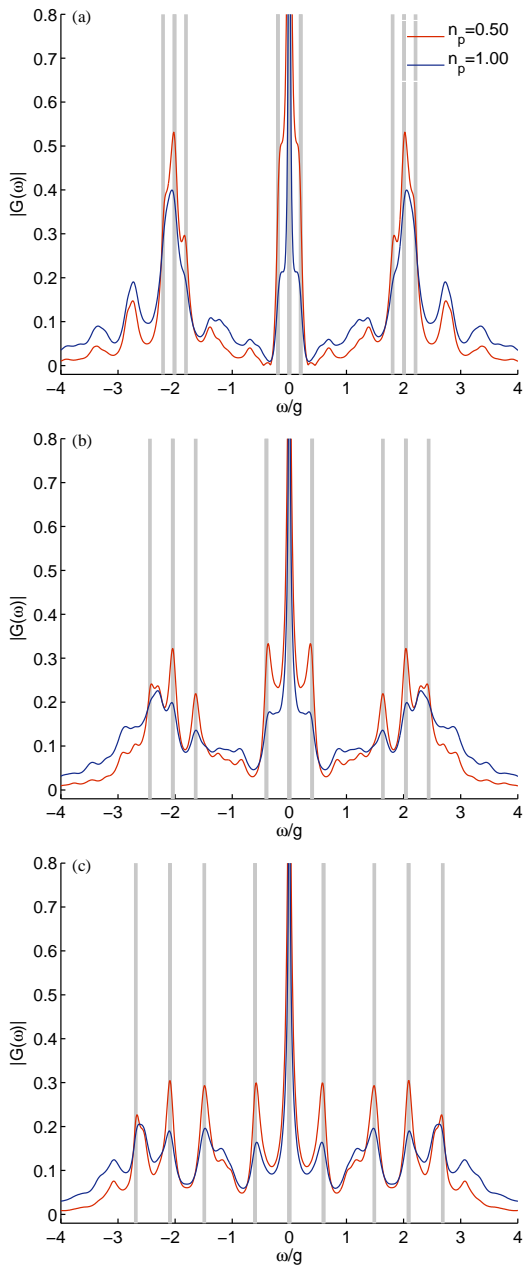}
        \caption{(Color online) Spectrum $G(\omega)$ of the second-order correlation function, for zero detuning $\Delta/g=0$, loss rate $\Gamma^-/g=0.2$ and hopping strength \fc{a} $J/g=0.2$, \fc{b} $J/g=0.4$, and \fc{c} $J/g=0.6$, respectively. The driving strength $\Gamma^+$ is adjusted such that the steady-state particle density is $n_p=\lbrace 0.5,\,1.0 \rbrace$. Light-gray (vertical) lines indicate \tra{1}{1}-transitions within the $\np=1$ sector of the non-dissipative two cavity system.}
        \label{fig:g2}
\end{figure}
\begin{figure*}
        \centering
        \includegraphics[width=0.9\textwidth]{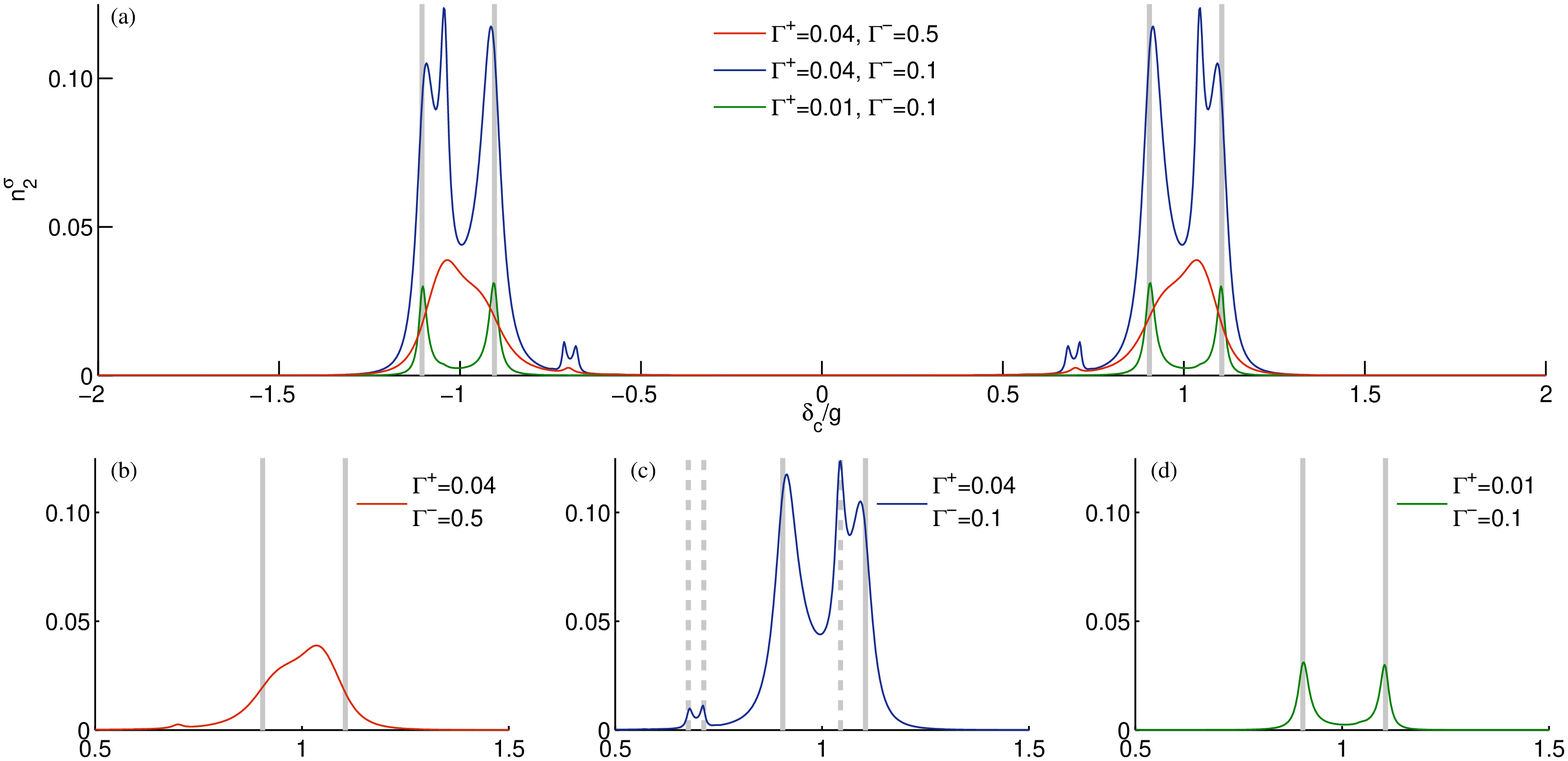}
        \caption{(Color online) Steady-state occupation $n_2^\sigma$ of the {\tls} in the second cavity, main panel \fc{a}, as a function of the coherent driving laser detuning $\delta_c$ for hopping strength $J/g=0.2$ and distinct values of the driving strength $\Gamma^+/g$ and loss rate $\Gamma^-/g$, see legend. Panels \fc{b}--\fc{d} magnify the occupation $n_2^\sigma$ for the positive detuning response of the individual rates. Solid, light-gray (vertical) lines indicate the eigenenergies $E_\nu^1$ of the $\np=1$ particle number sector of the non-dissipative two cavity system, whereas dashed, light-gray (vertical) lines in panel \fc{c} correspond to half of the eigenenergies $E_\mu^2/2$ of the $\np=2$ particle number sector; see main text for details. }
        \label{fig:nOfDetCoh}
\end{figure*}
Results are evaluated for hopping strength \fc{a} $J/g=0.2$, \fc{b} $J/g=0.4$, and \fc{c} $J/g=0.6$.  

Due to the similarity of  the second-order correlation function, given in \eq{eq:g2}, with  the density-density correlation function 
$\langle {\hat n}^\sigma_{2}(t) {\hat n}^\sigma_{2}(0)\rangle$, where ${\hat n}^\sigma_{2}=\sigma^+_2 \sigma^-_2$, one is prompted to analyze the results of the second-order correlation function in terms of the density-density correlation in the non-dissipative system. 
The density-density correlation  correspond to excitations within a Hilbert space sector of fixed particle number $\np$ and therefore probes \tra{\np}{\np}-transitions. 
The corresponding resonances are located at $\bar \omega_{\mu\nu}^\np = E_\mu^\np - E_\nu^\np$. Light-gray (vertical) lines in \fig{fig:g2} indicate the position of the poles originating from \tra{1}{1}-transitions of the non-dissipative two-cavity system $\bar \omega_{\mu\nu}^1 = E_\mu^1 - E_\nu^1$ as obtained from \eq{eq:en}. The spectrum $G(\omega)$ exhibits resonances at the poles $\bar \omega_{\mu\nu}^1$ along with a background that stems from density-density excitations $\bar \omega_{\mu\nu}^\np$ for the Hilbert space sectors with larger particle number $\np>1$.

\section{\label{sec:coherent} Coherent driving}
Here, we investigate coherent driving on the {\tls} modeled with the additional Hamiltonian term introduced in \eq{eq:coh}. In this section, we restrict our analysis to zero detuning between the {\tls} and the cavity photons, \ie, $\omega_c=\epsilon$ and thus $\Delta/g=0$. From this restriction it follows that the detuning $\delta_\epsilon$ between the driving laser and the {\tls} is equal to the detuning $\delta_{\omega_c}$ between the driving laser and the cavity photons. In the following, we denote this coherent driving detuning as $\delta_c\equiv \delta_\epsilon=\delta_{\omega_c}$. 

The coherently driven system is populated only when the driving laser is in resonance with a transition of the two-cavity system. For vanishing loss rate and very weak driving only the $\np=1$ sector of the initially empty two-cavity system can be occupied, since all transitions involving higher particle numbers are detuned from the \tra{0}{1}-transitions and are thus prohibited. However, the dissipation is responsible for an energy level broadening as compared to the non-dissipative system shown in  \figc{fig:ladder}{b} and therefore has a drastic impact on the occupation. Following these considerations, the energy spectrum of the coherently driven two-cavity system can be probed by measuring the occupation of the {\tls} in the second cavity $n_2^\sigma$ as a function of the coherent driving laser detuning $\delta_c$, which is accessible in experiments by measuring the output field \cite{wallraff_strong_2004,johnson_quantum_2010}.

Numerical results for the steady-state occupation $n_2^\sigma$ as a function of the coherent driving detuning $\delta_c$ are depicted in \fig{fig:nOfDetCoh}, main panel \fc{a}, for hopping strength $J/g=0.2$, and rates $\lbrace \Gamma^+/g=0.04$, $\Gamma^-/g=0.5\rbrace$, $\lbrace\Gamma^+/g=0.04$, $\Gamma^-/g=0.1\rbrace$, and  $\lbrace\Gamma^+/g=0.01$, $\Gamma^-/g=0.1\rbrace$, see \eq{eq:coh}. A magnification of the positive detuning response for the respective rates are shown in panels \fc{b}--\fc{d}. For large loss rate $\Gamma^-=0.5$ and driving $\Gamma^+=0.04$, see panel \fc{b}, the two peak structure expected for \tra01-transitions cannot be resolved due to the level broadening and only one peak at $|\omega-\omega_c|\approx g$ is observed in the occupation spectrum. When decreasing the loss rate to $\Gamma^-=0.1$, and keeping the driving fixed at $\Gamma^+=0.04$, see panel \fc{c}, the expected four peak structure emerges. However, apart from the \tra01-transitions, which are indicated by solid light-gray (vertical) lines, additional resonances at  $|\omega-\omega_c|\approx 0.7g$ and $|\omega-\omega_c|\approx g$ are present. These resonances are identical to half of the eigenenergies $E^2_\mu$ of the $\np=2$ particle number sector and originate from subsequent \tra01- and \tra12-transitions. The energies  $E^2_\mu/2$ are indicated by dashed light-gray (vertical) lines in panel \fc{c}. In the non-dissipative case these transitions are not allowed because each \tra01-transition is detuned from the \tra12-transitions. However, for large enough driving strength and for finite loss rate, \ie, finite level broadening, these transitions become possible. The additional peaks, however, disappear when reducing the driving strength to $\Gamma^+=0.01$, as shown in panel  \fc{d}, where the only remaining peaks stem from the \tra01-transitions. 

The fluorescence spectrum $S(\omega)$ of a coherently driven system is non-zero only when the driving laser is in resonance with the two-cavity system, meaning that the driving laser is able to populate the two-cavity system. Importantly, a resonance peak in the fluorescence spectrum is observed at the specific frequency imprinted from the detuning of the coherent driving laser $\delta_c$. The only exception takes place for additional transitions that emerge due to the level broadening.

\section{\label{sec:conclusion}Conclusions}

In the present paper, we investigated a laser-driven and dissipative system of two coupled cavities with Jaynes-Cummings nonlinearity. In particular, we presented and discussed the fluorescence spectrum $S(\omega)$ and the spectrum $G(\omega)$ of the second-order correlation function $g^{(2)}(t)$ evaluated in steady state, which is of utmost relevance for experiments. All quantities are obtained by solving the time dependence of the full quantum mechanical master equation numerically by means of the Arnoldi time evolution method introduced here to treat open quantum systems. 

Each cavity of the investigated system is described by the Jaynes-Cummings model and thus confines photons and contains a two-level system. The two-level system couples to the photons, which gives rise to the nonlinearity. The coupling between the two cavities results from a finite overlap of respective photonic wave functions. In order to incorporate dissipation and driving we employed a master equation in Lindblad form. Specifically, we consider two experimentally distinct kinds of driving on the two-level system of the first cavity: (i) \textit{incoherent} driving, which takes advantage of a three level structure, that is modeled by a Lindbladian pump term in the master equation approach and (ii) \textit{coherent} driving which is modeled by an additional Hamiltonian term. The dissipation is taken into account by a Lindbladian loss term coupling to the two-level system of the second cavity. 

In this work, we focused on relating the resonance peaks of the measured spectra to the energy spectra of the non-dissipative two-cavity system. In the case of incoherent driving, resonance peaks in the fluorescence spectra could be related to excitations from the $N_p$- to $(N_p+1)$-particle sector of the non-dissipative two-cavity system. In the weak driving limit the main contributions emerge from excitations corresponding to the $N_p=0$ to $N_p=1$ transitions. For increasing driving strength, however, excitations from higher particle number sectors, such as $\np=1$ to $\np=2$ transitions become more pronounced. In order to gain insight into the measured spectra, we also presented an approximate prescription to calculate the spectrum semianalytically from the steady-state density matrix and from the excitation energies of the non-dissipative system. Finally, we demonstrated that the spectrum $G(\omega)$ of the second-order correlation function probes excitations within a certain particle number sector. Therefore it is related to the density-density correlations. For coherent driving the driving laser has to be tuned to be in resonance with a transition of the two-cavity system.  This is necessary to populate the system.  Thus we mapped out the energy spectrum of the non-dissipative two-cavity system by measuring the occupation of the two-level system in the second cavity. In the case of resonance, the spectrum yields a response at the detuning between the driving laser and the system, otherwise the spectrum is virtually zero. Additional peaks may occur due to the finite level broadening arising from the photon loss term. In summary, for coherent driving, the occupation number provides a simple approach to obtain the excitation energies of the non-dissipative two-cavity system for small filling factors, whereas for incoherent driving, the spectra yield the full information about the excitation energies for arbitrary filling factors.

\begin{acknowledgments}
We thank A.~D.~Greentree for fruitful discussions.
We made use of parts of the ALPS library \cite{albuquerque_alps_2007}.
We acknowledge financial support from the Austrian Science Fund (FWF)
under the doctoral program ``Numerical Simulations in Technical
Sciences'' Grant No. W1208-N18 (M.K.) and under Project No. P18551-N16
(E.A.).
\end{acknowledgments}

%

\end{document}